\documentclass[reprint,amsmath,amssymb,aps,nofootinbib,noeprint]{revtex4-2}

\usepackage{epsfig}
\usepackage{verbatim}
\usepackage{epstopdf}
\usepackage{graphicx}
\usepackage{aas_macros}
\usepackage{epsf}
\usepackage{amsmath}
\usepackage{xcolor}
\usepackage{tikz}
\usepackage{enumitem}
\usepackage{adjustbox}
\usepackage{threeparttable}
\usepackage{ragged2e}

\definecolor{darkgreen}{rgb}{0.0,0.5,0.0}
\usepackage[colorlinks,citecolor=darkgreen]{hyperref}

\usepackage[leftmargin=0pt,rightmargin=0pt,skipabove=2pt,skipbelow=2pt,innerleftmargin=5pt,innerrightmargin=5pt,innertopmargin=5pt,innerbottommargin=5pt,linewidth=0pt,innerlinewidth=0pt,middlelinewidth=0pt,outerlinewidth=0pt,roundcorner=0pt]{mdframed}

\newcommand{\ie}{\emph{i.e.} }

\newcommand{\be}{\begin{equation}}
\newcommand{\ee}{\end{equation}}
\newcommand{\bea}{\begin{equation*}}
\newcommand{\eea}{\end{equation*}}
\newcommand{\beqr}{\begin{eqnarray} \nonumber}
\newcommand{\eeqr}{\end{eqnarray}}
\newcommand{\beqrb}{\begin{eqnarray}}
\newcommand{\eeqrb}{\nonumber \end{eqnarray}}
\newcommand{\fin}{\mbox{ .}}
\newcommand{\coma}{\mbox{ ,}}

\newcommand{\Max}{{\mathrm{max}}}

\newcommand{\cm}{\mbox{ cm}}
\newcommand{\sr}{\mbox{ sr}}
\newcommand{\se}{\mbox{ s}}

\newcommand{\erg}{\mbox{ erg}}

\newcommand{\MHz}{\mbox{ MHz}}
\newcommand{\GHz}{\mbox{ GHz}}

\newcommand{\pc}{\mbox{ pc}}
\newcommand{\kpc}{\mbox{ kpc}}

\newcommand{\eV}{\mbox{ eV}}
\newcommand{\keV}{\mbox{ keV}}

\newcommand{\GeV}{\mbox{ GeV}}
\newcommand{\TeV}{\mbox{ TeV}}
\newcommand{\PeV}{\mbox{ PeV}}

\newcommand{\muG}{\mbox{ $\mu$G}}

\newcommand{\gama}{$\gamma$}

\newcommand{\vect}[1]{\mathbf{#1}}
\newcommand{\unit}[1]{\hat{\mathbf{#1}}}

\newcommand{\mynewcommand}[2]{\ifdefined #1 \else \newcommand{#1}{#2} \fi}

\mynewcommand{\apj}{ApJ}     
\mynewcommand{\apjl}{ApJL}     
\mynewcommand{\apjs}{ApJS}    
\mynewcommand{\aap}{A\&A}    
\mynewcommand{\nat}{Nature}  

\newcommand{\dgr}{^{\circ}}

\newcommand{\dgrdot}{{\overset{^\circ}{.}}}

\newcommand{\Mach}{\mathcal{M}}
\newcommand{\myEps}{{\varepsilon}}

\setcitestyle{super}

\makeatletter
\def\@hangfrom@section#1#2#3{\@hangfrom{#1#2}#3}
\def\@hangfroms@section#1#2{#1#2}
\makeatother

\begin{document}

\title{Fermi bubbles detected in $\sim$100 TeV neutrinos}

\author{Uri Keshet}
\address{
    Physics Department, Ben-Gurion University of the Negev, POB 653, Be'er-Sheva 84105, Israel
}
\email{keshet.uri@gmail.com}

\author{Ilya Gurwich}
\address{
    Qedma Quantum Computing, Tel-Aviv, Israel
}

\begin{abstract}
\begin{mdframed}[backgroundcolor=black!5]
The Fermi bubbles\cite{DoblerEtAl10,SuEtAl10} have been identified as a collimated\cite{MondalEtAl22} bipolar outflow emanating from the Milky-Way center, tracing strong forward shocks\cite{KeshetGurwich17,KeshetGurwich18} which extend $\sim10$ kpc from the Galactic disk, originating from a $\sim10^{56\pm1}$ erg outburst\cite{GhoshEtAl26}.
These shocks are sufficiently strong, extended, and energetic to produce a detectable flux of $\gtrsim 10\TeV$ neutrinos, especially in their denser north\cite{KeshetGhosh26} and east\cite{GhoshEtAl26}, but early IceCube data were insufficient for identifying the signal\cite{SherfEtAl17,FangEtAl17}.
We find that IceCube high-energy starting events (HESE 12-year\cite{ChirkinEtAl24}) correlate with the $\textit{Fermi}$-LAT sky map outside the Galactic plane ($>3\sigma$).
Testing for neutrinos coincident with the bubble shells, localized using $\textit{eROSITA}$ data\cite{PredehlEtAl20}, we detect ($>4\sigma$) both bubbles at high ($|b|>30^{\circ}$) latitudes, with a local excess mainly in their X-ray bright eastern shells ($>5\sigma$).
The signal matches the anticipated secondaries of relativistic ions carrying $\sim 10^{54.5\pm0.5}$ erg in each bubble, shock-accelerated to $>$PeV energies.
The results verify the strength of the shocks, suggesting an ion acceleration efficiency of order $\sim10\%$.
We also present preliminary evidence for neutrinos from the even larger shells of the eROSITA bubbles, which encapsulate their younger Fermi-bubble counterparts, carrying a similar energy\cite{GhoshEtAl26} and confined by shocks nearly as strong\cite{KeshetGhosh26}.
\end{mdframed}
\end{abstract}

\maketitle

\section{Introduction}
\label{sec:Intro}

The \emph{Fermi} bubbles (FBs) are the manifestation of an extended bipolar outflow emanating from the Milky Way's Galactic center (GC)\cite{Baganoffetal03, Blandhawthorncohen03}, rising $\sim10\kpc$ nearly perpendicular to the Galactic plane, where they are traced in inverse-Compton \gama-rays\cite{DoblerEtAl10,SuEtAl10}, thermal X-rays\cite{KeshetGurwich18}, and polarized\cite{Keshet25PolFB} and non-polarized\cite{KeshetEtAl24} synchrotron radio and dust microwave emission.
Their combined, approximately uniform non-thermal signature\cite{SuEtAl10, Dobler12, HooperSlatyer13, HuangEtAl13, PlanckHaze13, Ackermannetal14} is naturally modeled, without introducing spurious energy cutoffs, as leptonic emission from cosmic-ray electrons (CREs)\cite{KeshetEtAl24}, accelerated to $\gtrsim$TeV energies by the strong, $\Mach\gtrsim 5$ shocks\cite{KeshetGurwich17, KeshetGurwich18} at their expanding edges, as they gyrate in the shock-amplified\cite{Keshet25PolFB} magnetic fields, Compton-scatter optical (primarily\cite{KeshetEtAl24}) photons, and diffuse downstream\cite{KeshetGurwich17}. Neutrinos, expected from cosmic-ray ions (CRIs) accelerated by these extended, strong shocks to $\sim10^{17}\eV$ energies, have until now remained undetected\cite{SherfEtAl17,FangEtAl17,AbbasiEtAl23}, but their $>100\TeV$ neutrino signature was expected to emerge when IceCube\cite{IceCube17} collects sufficient data.

Meanwhile, the FBs and their environment have become better understood, in particular as \emph{eROSITA} X-rays\cite{PredehlEtAl20} have verified earlier radio\cite{Sofue77} and X-ray\cite{Sofue94} indications, especially from \textit{ROSAT}\cite{Sofue00}, for an even larger, older Galactic pair of so-called ROSAT/eROSITA bubbles (RBs).
As the southern RB finally emerged clearly in X-rays\cite{PredehlEtAl20}, facilitating the detection of its edges in radio and \gama-rays\cite{KeshetGhosh26}, the bipolar nature of its Loop-I northern counterpart became established.
While the RBs extend out to extreme, $|b|\sim 80\dgr$ Galactic latitudes, encapsulating (not only in projection\cite{GhoshEtAl26}) the $|b|\lesssim 50\dgr$ FBs, they too show strong shocks at their edges\cite{KeshetGhosh26} and likely arise from a very similar collimated GC outburst of comparable $\sim 10^{56\pm1}\erg$ energy\cite{GhoshEtAl26}, challenging alternative suggestions\cite{Sarkar24, ZhangEtAl24}, and indicating that the FBs expand into a recently shocked circumgalactic medium (CGM)\cite{ZhangEtAl25,GhoshEtAl26}.

The northern RB is noticeably brighter than its southern counterpart, especially in its eastern sector\cite{Haslam82,Sofue00}, while the FBs are more comparable to each other in terms of integrated brightness\cite{SuEtAl10}.
However, near-shock downstream emission indicates a similar brightening in the north and east, not only in the RBs\cite{KeshetGhosh26}, but also in the FBs\cite{KeshetGurwich17,KeshetGurwich18,KeshetEtAl24}, suggesting an ambient density higher by a factor of $\sim2$ in the northern high-latitude hemisphere\cite{KeshetGhosh26}, with an eastern gradient\cite{GhoshEtAl26}.
Hence, searches for subtle, in particular hadronic signals from the Galactic bubbles should focus on the northern hemisphere and on the eastern sectors.
Such an updated search for IceCube neutrinos from the FBs is timely: while the four-year IceCube data release\cite{Aartsen2014, KopperEtAl15} indicated an upper $U_i\lesssim 10^{55.5}\erg$ limit\cite{SherfEtAl17} on the total CRI energy in these bubbles, present constraints suggest a total $U>10^{55}\erg$ in the FBs; a similar limit is also inferred in the RBs\cite{GhoshEtAl26}.

A high-latitude IceCube detection of the FBs is also anticipated by extrapolating the neutrino counterpart of their $\varepsilon I_\varepsilon(\varepsilon=10\varepsilon_{1}\GeV)\simeq 0.5\keV\se^{-1}\cm^{-2}\sr^{-1}$ leptonic \gama-rays\footnote{We denote neutrino energy by $\epsilon$ and photon energy by $\varepsilon$.}.
Hadronic \gama-rays expected from the FBs are fainter than their $\nu I_\nu\simeq 5 \varepsilon_{1}^{-0.7}\keV\se^{-1}\cm^{-2}\sr^{-1}$ inner Galactic-disk counterpart\cite{AckermannEtAl12} by a factor $\mathcal{R}\simeq 10^{3}\eta_{-3}\rho_{-3}^{-1}E_{2}^{-0.7}$.
Here, $\eta\equiv 10^{-3}\eta_{-3}$ is the CRE-to-CRI energy ratio, assumed constant at such $E\gg 1\GeV$ CRI energies, $\langle\rho\rangle\equiv\rho_{-3}10^{-3}m_p \cm^{-3}$ is the mean mass density averaged over the $|b|>30\dgr$ FB volume, $m_p$ is the proton mass, $E\equiv 100E_{2}\GeV$ is the energy of the parent CRI, and we follow earlier definitions of the inner Galaxy as $|b|<8\dgr$ and $|l|<80\dgr$, comparable in solid angle to the FBs.
The recent detection\cite{AbbasiEtAl23} of the inner Galactic-disk in $\myEps\gtrsim60\TeV$ neutrinos, where spectrally-flat FBs extrapolate to $\mathcal{R}\simeq 1.6\eta_{-3}/\rho_{-3}$, suggests an easily detectable FB signal, outshining the Galactic disk for $\eta<6\times 10^{-4}\rho_{-3}$.

\section{High-latitude neutrino--\gama-ray correlations}
\label{sec:Corr}

It is interesting to compare the high-latitude sky distributions of structure tracers vs. high-energy neutrinos. For the latter, we adopt the 12-year high-energy starting event (HESE\cite{ChirkinEtAl24}) all-sky sample distribution\footnote{See \url{https://icecube.wisc.edu/data-releases/2023/07/icecube-hese-12-year-data-release/}} $p(\unit{n})\equiv \sum_j p_j(\unit{n})$, summing the $p_j(\unit{n})$ sky probability density functions (PDFs) of events $j$ within a given $(\epsilon_{1},\epsilon_2)$ neutrino energy range.
The sky correlation is computed in a cylindrical equal-area $(l,\sin b)$ Galactic-coordinate projection, after masking the $|b|<30\dgr$ Galactic disk and stitching together the remaining map segments at the poles and across the masked disk by introducing a periodic coordinate $\sin b_*$.
The Pearson $r_p$ coefficient and normalized cross correlation $r_c$ are computed for each tracer map multiplied by the IceCube effective area $A(\unit{n})$, approximated as the all-flavor average based on the instrument response functions (IRFs) of the $7.5$-year IceCube data release \cite{AbbasiEtAl21}, flux-weighted within $(\epsilon_{1},\epsilon_2)$ for the expected $\propto \epsilon^{-2}$ spectrum.
In order to determine the significance of the correlation, the tracer map is pre-rotated by arbitrary $\Delta l$ and $\Delta \sin b_*$ shifts.
The resulting empirical null distributions of $r_p$ and $r_c$ are then used to estimate the p-value and corresponding $Z$ score of the real, \ie non-rotated, tracer--IceCube correlation.

\begin{figure}[h!]
    \includegraphics[width=0.49\textwidth,trim={0cm 0cm 0cm 0cm},clip]{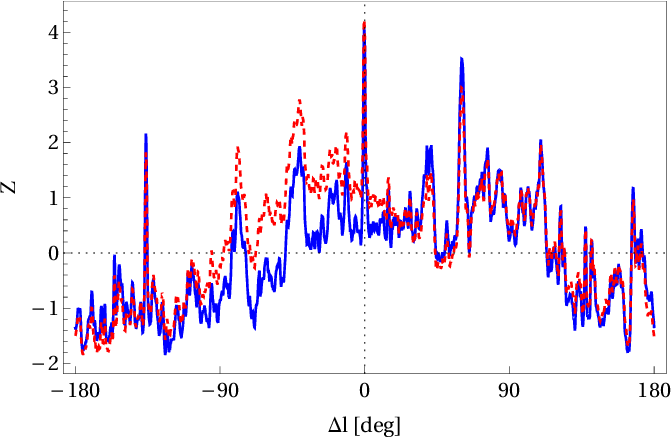}\\
    \vspace{0.3cm}
    \includegraphics[width=0.49\textwidth,trim={0cm 0cm 0cm 0cm},clip]{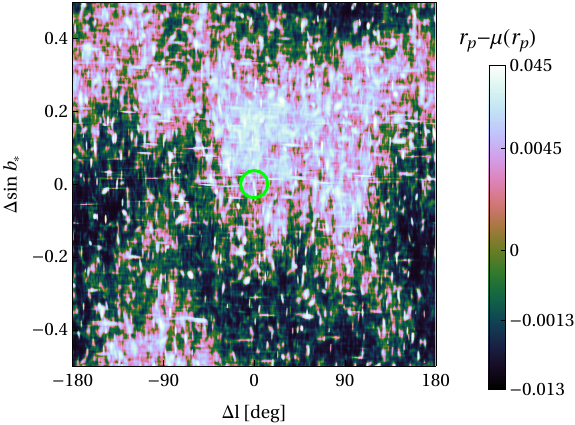}
	\caption{\label{fig:LAT4Corr}
         Correlation of the HESE 12-year PDF sky map with \emph{Fermi}-LAT $3$--$10\GeV$, rotated at different $\Delta l$ (upper panel shows $Z$ scores of $r_p$ as solid blue and $r_c$ as dashed red) as well as $\Delta\sin b_*$ ($r_p$ in bottom panel) and then weighed by $A$.
    }
\end{figure}

The results are summarized in Table \ref{tab:corr}, with Fig.~\ref{fig:LAT4Corr} demonstrating $\Delta l$ (top panel) and combined $\Delta l$ and $\Delta\sin b_*$ (bottom) rotations for the $3$--$10\GeV$ \emph{Fermi}-LAT (see \S\ref{sec:FermiLAT}) tracer.
The table and bottom panel of Fig.~\ref{fig:LAT4Corr} focus on $r_p$; slightly more significant \gama-ray correlations are obtained for $r_c$.
The results are well converged with (shown for nominal $\sim14'$) resolution, as demonstrated in \S\ref{app:CorrResolution}.

The correlations are positive for all tracers compared with the full, \ie $(10\TeV,10\PeV)$ HESE sample, and in most cases also with a high-energy, $\epsilon>100\TeV$ HESE sub-sample, although the latter has insufficient statistics for a meaningful conclusion.
Significant correlations are found between the full HESE sample and the \gama-ray tracers, in particular the $1$--$3\GeV$ ($3.4\sigma$ correlation) and $3$--$10\GeV$ ($3.7\sigma$) \emph{Fermi}-LAT channels, in which the FBs and RBs are most noticeable.
The FBs, and to a lesser degree also the RBs, are natural suspects behind these correlations, and are considered next.

\begin{table}[h!]
\begin{threeparttable}
\caption{\label{tab:corr} Photon--neutrino $|b|>30\dgr$ sky correlations}
\begin{tabular}{lrr}
\hline
Tracer photons\quad\quad\quad\quad & $\epsilon>10\TeV$ & \quad\quad\quad$\epsilon>100\TeV$ \\
\hline
$408\MHz$ & $0.068$ ($1.3\sigma$) & $0.061$ ($0.4\sigma$) \\
$30\GHz$ & $0.062$ ($0.9\sigma$) & $0.012$ ($-1.8\sigma$) \\
$44\GHz$ & $0.062$ ($0.7\sigma$) & $0.007$ ($-2.0\sigma$) \\
$70\GHz$ & $0.060$ ($0.5\sigma$) & $0.007$ ($-1.7\sigma$) \\
$0.3$--$0.6\keV$ & $0.067$ ($1.0\sigma$) & $0.126$ ($1.4\sigma$) \\
$0.6$--$1.0\keV$ & $0.066$ ($1.3\sigma$) & $0.096$ ($1.0\sigma$) \\
$1.0$--$2.3\keV$ & $0.056$ ($1.6\sigma$) & $0.066$ ($0.8\sigma$) \\
$0.3$--$1\GeV$ & $0.056$ ($1.9\sigma$) & $0.044$ ($0.2\sigma$) \\
$1$--$3\GeV$ & $0.050$ ($3.4\sigma$) & $0.031$ ($0.2\sigma$) \\
$3$--$10\GeV$ & $0.034$ ($3.7\sigma$) & $0.023$ ($0.4\sigma$) \\
$10$--$30\GeV$ & $0.018$ ($2.4\sigma$) & $0.016$ ($0.6\sigma$) \\
\hline
\end{tabular}
\par\footnotesize
\setlength{\emergencystretch}{2em}
\justifying
\noindent Pearson coefficient $r_p$ (with $Z$ score in parenthesis) for sky tracers of different radio\cite{Haslam82}, microwave\cite{Planck20LFI}, X-ray (all availble\cite{PredehlEtAl20} all-sky channels), and \gama-ray photon regimes (rows), correlated with HESE 12-year neutrinos across two energy thresholds (columns). \\
\end{threeparttable}
\end{table}

\begin{figure*}
    \includegraphics[width=0.97\textwidth,trim={0.05cm 0.5cm 0.05cm 0.5cm},clip]{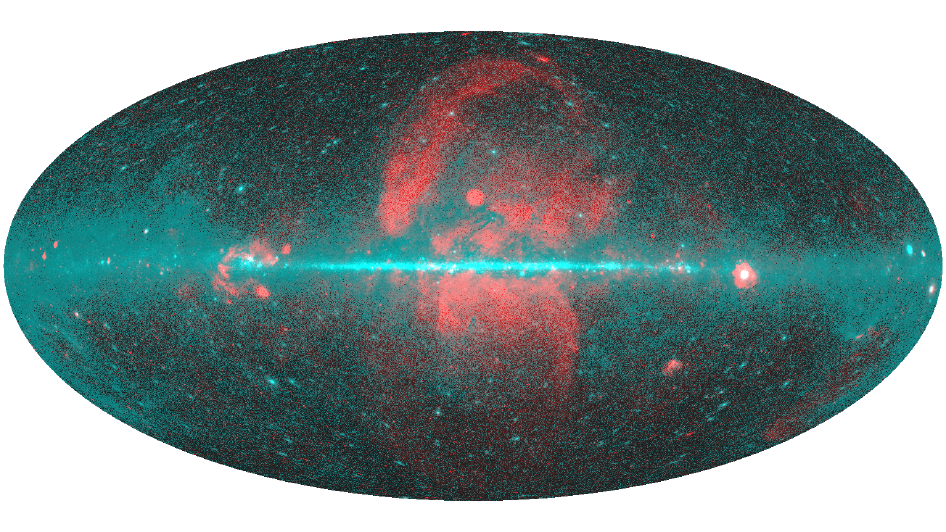}
	\caption{\label{fig:eROSITA2}
    Galactic-coordinate Hammer-Aitoff projection of \emph{Fermi}-LAT $1$--$3\GeV$ (cyan) and \emph{eROSITA} $0.6$--$1.0\keV$ (red).
   	}
\end{figure*}

\begin{figure*}
    \includegraphics[width=0.97\textwidth,trim={0.cm 0.2cm 1.0cm 0.2cm},clip]{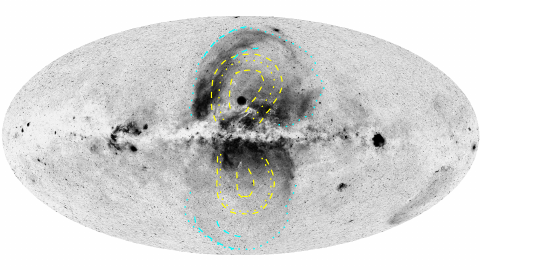}
	\caption{\label{fig:eROSITA2Shells}
    \emph{eROSITA} $0.6$--$1.0\keV$ image with superimposed bubble features.
    FB shells are roughly traced by edge-detected steepest \emph{Fermi}-LAT gradients\cite{KeshetGurwich17} (dotted yellow contours) and demarked (dashed yellow) according to \emph{ROSAT}\cite{KeshetGurwich18} and \emph{eROSITA} profiles (see Fig.~\ref{fig:eRositaProf}).
    RB shells are traced by edge-detected steepest \emph{eROSITA} gradients (dotted cyan) most robustly in the high-latitude eastern sectors\cite{KeshetGhosh26}, where they are demarked (dashed cyan) by \emph{eROSITA} profiles (see Fig.~\ref{fig:RBs}).
   	}
\end{figure*}

\section{Fermi and eROSITA bubbles and shells}
\label{sec:Xrays}

The bipolar Galactic bubbles are evident in the $0.6$--$1.0\keV$ \emph{eROSITA}\cite{PredehlEtAl20} and $1$--$3\GeV$ \emph{Fermi}-LAT images combined in Fig.~\ref{fig:eROSITA2}, with bubble features superimposed on the \emph{eROSITA} map in Fig.~\ref{fig:eROSITA2Shells}.
Both RBs and FBs are strongly limb-brightened in their northeast sectors, suggesting a locally elevated ambient density compressed by the two consecutive shocks.
Some limb brightening is seen across the entire RB X-ray shell, facilitating its full coarse-grained edge-detection (dotted cyan curve), although only the high-latitude eastern sectors are robustly localized\cite{KeshetGhosh26}.
The FBs are similarly limb brightened in X-rays, but the effect is easily discernable only at low latitudes; the high-latitude shells are better edge-detected (dotted yellow) in \gama-ray \emph{Fermi}-LAT images\cite{KeshetGurwich17} before being substantiated in X-rays\cite{KeshetGurwich18}.

\begin{figure}[h]
    \begin{tikzpicture}
    \draw (0, 0) node[inner sep=0, align=center]
    {
        \includegraphics[width=0.47\textwidth,trim={0cm 0cm 0cm 0cm},clip]{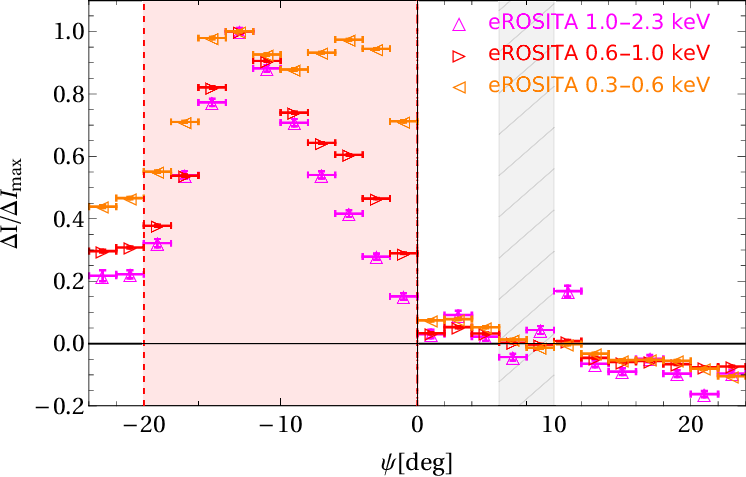}\\[1ex]
        \includegraphics[width=0.47\textwidth,trim={0cm 0cm 0cm 0cm},clip]{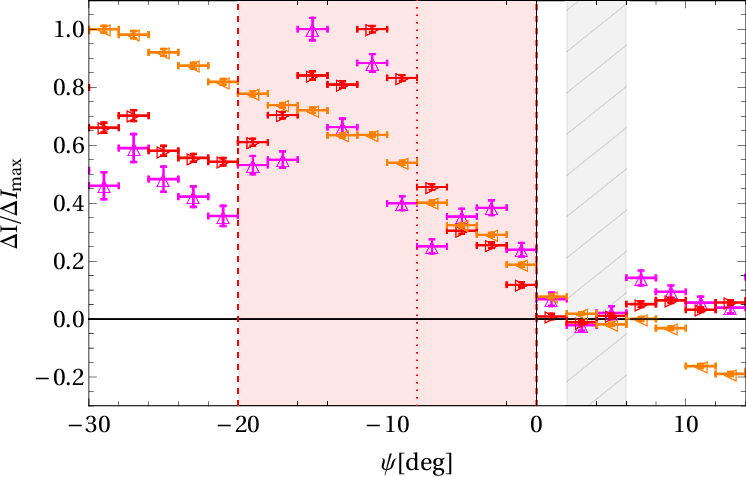}
    };
    \end{tikzpicture}
	\caption{\label{fig:RBs}
        Stacked \emph{eROSITA} brightness profiles in bins parallel to the northeast (top panel) and southeast (bottom) RB edges\cite{KeshetGhosh26}, normalized to $\Delta I=0$ upstream (at a selected flat, gray hatched region), in three energy channels (top-panel legend).
        The steepest \emph{eROSITA} gradient (in the $0.6$--$1.0\keV$ channel; vertical dotted red line) coincides with the inferred $\psi=0$ shock in the north, but lies $\Delta\psi=-8^\circ$ downstream of the shock in the south.
        The limb-brightened shell is demarked (red shaded between vertical red dashed lines).
        Error bars designate the dispersion within each bin (henceforth).
    }
\end{figure}

\begin{figure}[h]
    \includegraphics[width=0.47\textwidth,trim={0cm 0cm 0cm 0cm},clip]{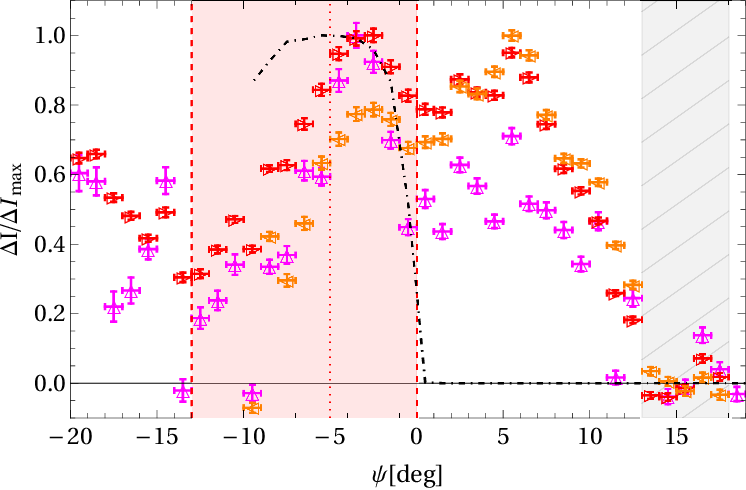}\\
    \includegraphics[width=0.47\textwidth,trim={0cm 0cm 0cm 0cm},clip]{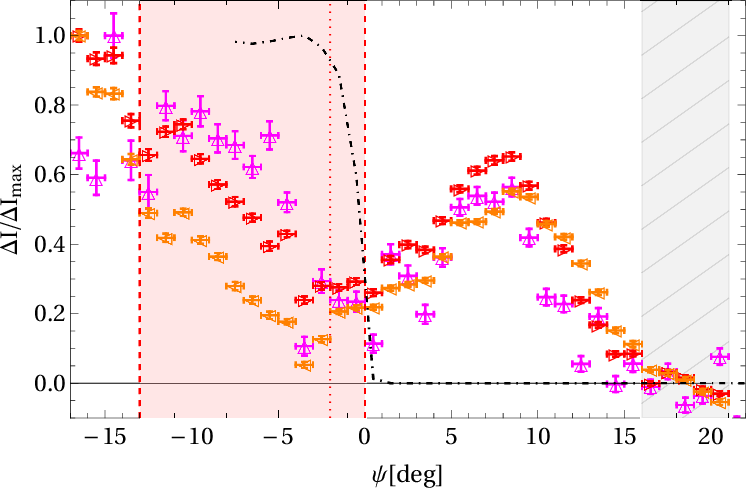}
	\caption{\label{fig:eRositaProf}
        Same as Fig.~\ref{fig:RBs} but for the north (top) and south (bottom) $|b|>30\dgr$ FB edges. The \emph{eROSITA} brightness profiles
        substantiate the \emph{ROSAT}\cite{KeshetGurwich18} shell localization (red shaded between dashed lines); the coarse-grained steepest \emph{Fermi}-LAT gradient\cite{KeshetGurwich17} (vertical dotted red line) lies in the north $\Delta\psi=-5^\circ$ (in the south $\Delta\psi=-2^\circ$) downstream of the implied $\psi=0$ shock.
    }
\end{figure}

The projected FB and RB shells can be traced by stacking data parallel to the edge-detected curves of Fig.~\ref{fig:eROSITA2Shells}, as illustrated in Fig.~\ref{fig:RBs} for the high-latitude eastern sectors of the RBs and in Fig.~\ref{fig:eRositaProf} for the $|b|>30\dgr$ FBs.
The RB upstream is fairly uniform in X-rays, sharply rising as one crosses inward of the implied shock\cite{KeshetGhosh26}, denoted $\psi=0$ such that the angular distance from the shock is $\psi>0$ upstream ($\psi<0$ downstream).
This shock coincides with the detected edge in the north, but the south edge detector picks up a stronger gradient found $\sim8\dgr$ downstream of this shock, suggesting some substructure.
The X-ray emission is seen to be limb-brightened within $\sim20\dgr$ from the shock in both north and south sectors.

In the FBs, the upstream is less pristine, possibly due to the effects of the preceding RBs, presenting X-ray substructure peaking $\sim10\dgr$ upstream of the detected edge.
However, the shock can still be localized as a break in the X-ray profile\cite{KeshetGurwich18}, lying $\sim6\dgr$ ($2\dgr$) outside the strongest \gama-ray gradient in the north (south) bubble.
The limb-brightened X-ray shells span $\sim 13\dgr$ from the shock, where a break in the X-ray profile is inferred and is more evident in the north.
Thus, as in the RBs, the shell appears to span $\sim1/4$ of the shock radius, consistent with the effective shell width of a strong shock expanding into an isothermal-like halo.

Studies of thermal and nonthermal emission from the bubbles suggest that hadronic emission should show some limb brightening, especially in the northeast sector.
The hadronic signature depends on the projected distributions of target gas and CRs of relevant energies, the latter being sensitive to diffusion and magnetic confinement in the shell.
The target nuclei are strongly limb brightened in projection, as indicated by the X-ray emission measure.
The projected distribution of $\lesssim100\GeV$ CRs is limb brightened too, according to the strong localized brightening of radio emission from the RBs\cite{KeshetGhosh26} and evidence for some limb brightening of \gama-ray emission from both FBs\cite{KeshetEtAl24} and RBs\cite{KeshetGhosh26}.
One also expects hadronic emission to strengthen at low latitudes, but this depends on the post-outburst replenishment of gas and magnetic fields near the disk and on the diffusion of CRs against these fields.
Note that FB synchrotron emission strengthens considerably inside $|b|<30\dgr$, suggesting strong magnetization\cite{Dobler12LastLook}.

\section{North Fermi bubble in HESE neutrinos}
\label{sec:NFB}

\begin{figure*}
    \includegraphics[width=0.97\textwidth,trim={0cm 0.2cm 1.05cm 0.2cm},clip]{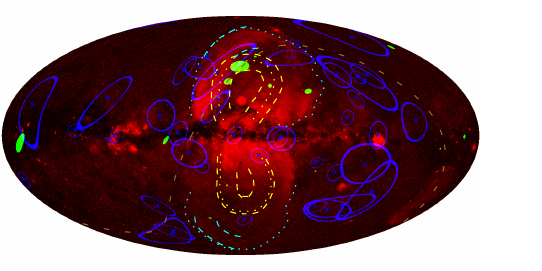}
	\caption{\label{fig:HESE100eROSITA}
    High, $>100\TeV$ energy IceCube neutrino tracks (14 green $>0.1p_{\Max}$ disks) and showers (41 events in highlighted $\sim0.5p_{\Max}$ blue contours around x marks) superimposed on the $1$--$3\GeV$ Fermi-LAT map in a Galactic-coordinate Hammer-Aitoff projection.
    FB shells are traced by edge-detected steepest \emph{Fermi}-LAT gradients\cite{KeshetGurwich17} (dotted yellow contours) and bounded (see Fig.~\ref{fig:eRositaProf}) by \emph{ROSAT}\cite{KeshetGurwich18} and \emph{eROSITA} X-rays (dashed yellow contours); RB edge-detected steepest \emph{eROSITA} gradients\cite{KeshetGhosh26} are also shown (in cyan; coarse grained as dot-dashed; fine grained as solid).
    The IceCube effective area at these energies diminishes north of the celestial equator (long-dashed brown).
   	}
\end{figure*}

\begin{figure*}
    \includegraphics[width=0.97\textwidth,trim={0.35cm 0.5cm 1.8cm 0.5cm},clip]{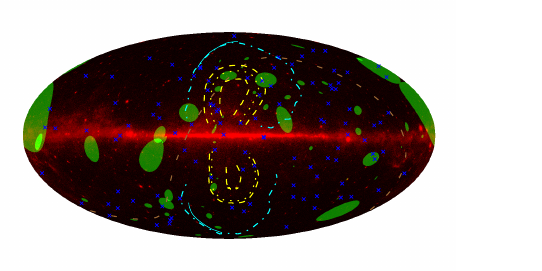}
	\caption{\label{fig:RGBAllx}
    Same as Fig.~\ref{fig:HESE100eROSITA}, but for all HESE events (showers shown only as blue x marks without ellipses to reduce clutter), superimposed on the $1$--$3\GeV$ \emph{Fermi}-LAT map.
   	}
\end{figure*}

The high-energy, $\epsilon > \epsilon_{min} = 100\TeV$ HESE 12-year events are highlighted in  Fig.~\ref{fig:HESE100eROSITA}, superimposed on the \emph{eROSITA} $0.6$--$1.0\keV$ map; all HESE events are similarly shown in Fig.~\ref{fig:RGBAllx} with no energy threshold, superimposed on the $1$--$3\GeV$ \emph{Fermi}-LAT map.

The choice of $\epsilon_{min}$ is motivated by atmospheric neutrinos and muon events, estimated to contribute $\sim 15$--$20\%$ of the $\epsilon>60\TeV$ HESE sample\cite{AbbasiEtAl21}, but only $\sim 6$--$8\%$ above $100\TeV$ for the expected hard, $dN/d\epsilon\propto\epsilon^{-2}$ spectrum of astrophysical neutrinos, giving $>50\%$ purity even in unvetoed track searches, as in previous work\cite{Aartsen2014, AartsenEtAl16}.
The 55 high-energy PDFs are highlighted in Fig.~\ref{fig:HESE100eROSITA}, coloring $p_j>0.1$ in green for the 14 tracks and $p_j\simeq 0.5$ in blue for the 41 showers, also designated by x-marks;
bubble edges and shells are superimposed as in Fig.~\ref{fig:eROSITA2Shells}.
In this section and in \S\ref{sec:NEShell} we consider the global neutrino signature of the northern FB, before the detailed modelling (\S\ref{sec:mod}) and analysis (\S\ref{sec:shells}) of the FB and RB shells and their local signatures.

The northern, $b>30\dgr$ FB subtends a solid angle $\Omega_0\sim0.20\sr$, or a fraction $\sim 1.6\%$ of the sky; weighing by $A$, this fraction becomes $\simeq 2.1\%$.
Hence, out of the 55 high, $>100\TeV$ energy HESE events, one expects about one event within this northern FB, whereas about seven such events are seen in Fig.~\ref{fig:HESE100eROSITA}, as listed in Table ~\ref{tab:nus}.
A random fluctuation producing such a high number of events within the northern FB is excluded at the $99.98\%$ ($3.6\sigma$) confidence level.
Focusing on the better-localized tracks, we find five events safely within the northern FB, out of 14 tracks across the entire sky, excluded as a fluctuation at a higher, $4.3\sigma$ confidence level.

Such global estimates for the significance of a FB neutrino signal, not accounting for priors, the precise overlap of the PDF with the bubble or its shell, and additional neutrino signals, are conservative.
For instance, using X-rays and radio as a prior, a signal is expected from the limb-brightened, thick FB shell, which spans a slightly smaller, $0.185\sr$ solid angle, corresponding to a $4.6\sigma$ track detection.
Removing events associated with the foreground, the Galactic-disk\cite{AbbasiEtAl23}, or (see \S\ref{sec:mod}) the south FB and the RBs would further raise the significance of the north FB within the remaining neutrinos.
For instance, if $5$--$10\%$ of the detected flux above 100 TeV is associated with the Galactic disk, as inferred from an earlier (non-HESE) IceCube analysis\cite{AbbasiEtAl23}, then this component should account for $1$--$2$ of the 14 tracks above 100 TeV; indeed, three out of these 14 tracks lie within $|b|<4\dgr$, greatly exceeding expectations for an isotropic model.
If a total of four of the 14 tracks are conservatively attributed to the disk, atmospheric foreground, the south FB, and the two RBs, the track signal of the northern FB would exceed $5\sigma$ despite the small-number statistics.

For our assumed flat, $dN_i/dE_i=C_i(\vect{r}) E_i^{-2}$ CRI spectrum, the conservative excess of $N=6$ neutrinos above $\epsilon_{1}=100\TeV$ corresponds to a luminosity
\begin{equation}
\!\!\!L(\epsilon_{1},\epsilon_2) \simeq \frac{4\pi \epsilon_{1}N d^2}{A\, t} \ln\left(\frac{\epsilon_{2}}{\epsilon_{1}}\right) \simeq 8.1 \times 10^{35} d^2_1 \erg\se^{-1}
\label{eq:Lnu}
\end{equation}
from the northern FB at $b>30\dgr$.
Here, $d=10d_{1}\kpc$ is the distance to the center of this FB, we adopted an exposure time $t\simeq11.4$ ($95\%$ duty cycle\cite{AbbasiEtAl21} for $12$) years\cite{ChirkinEtAl24}, and used a north $b>30\dgr$ bubble-averaged $A\simeq 18.1\mbox{ m}^2$.

We adopt a simple hadronic model, with an inelastic $\sigma_i\simeq 70\mbox{ mb}$ cross section for $\pi^\pm$ production at $\sim$PeV energies\cite{PDG2024}, resulting in a fraction $f_\nu\simeq0.15$ of the parent CRI energy evenly distributed on average among the arriving three neutrino flavors.
The luminosity \eqref{eq:Lnu} then indicates that the total CRI energy in the high-latitude north FB is
\begin{equation}\label{eq:ECRI}
  U_{i} \simeq \frac{\ln(\gamma_{\Max})}{\sigma_i c f_{\nu} \langle \rho\rangle/m_p} \frac{L(\epsilon_{1},\epsilon_{2})}{\ln\left(\epsilon_{2}/\epsilon_{1}\right)}
  \simeq 1.0\times10^{55}\frac{d_1^2}{\rho_{-3}}\erg \coma
\end{equation}
where we took a Hillas\cite{Hillas84} maximal $\gamma\simeq10^8$ CRI Lorentz factor.
If the northern FB has twice the CRI energy of its southern counterpart, then the total FB outburst energy deposited in CRI would exceed Eq.~\eqref{eq:ECRI} by $50\%$, or by a larger factor accounting for high-energy CRI reaching low, $|b|<30\dgr$ latitudes in either bubble.

The CRI energy budget \eqref{eq:ECRI}, even if multiplied by a factor of a few, is much too low to account for the observed $\gamma$-ray flux of the FBs in the $1$--$100\GeV$ range as $\pi^0$ decay, further validating the origin of these $\gamma$-rays as leptonic.
The energy of CREs in the FBs can be estimated from their radio-to-microwave `haze' emission, presenting an approximately constant $\nu^{1/2}I_\nu$ energy spectrum at $\nu\lesssim30\GHz$ frequencies below the cooling break\cite{KeshetEtAl24}, roughly comparable in both hemispheres.
The
$T_b(23\GHz)\simeq 30T_{30}\,\mu\mbox{K}$
brightness temperature averaged over $b<-35\dgr$ in the south FB\cite{Dobler12, PlanckHaze13}, if assumed to approximately hold symmetrically also in the $b>30\dgr$ north bubble, then provides the CRE counterpart of Eq.~\eqref{eq:ECRI},
\begin{eqnarray}\label{eq:ECRE}
  U_e & \simeq & A\frac{\left(m_e c\nu\right)^{5/2}}{(e^{7}B^{3})^{1/2}} \ln (\gamma_{e,\Max})  k_B T_b \Omega_0 d^2 \, \nonumber \\
  & \simeq & 5.9\times 10^{51}B_{0}^{-3/2} d_1^2 T_{30}\erg \fin
\end{eqnarray}
Here, $B=1 B_0\muG$ is the mean co-spatial magnetic field, the numerical constant\cite{RybickiBook79} $A\simeq3.2$, and $\gamma_{e,\Max}\simeq10^8$ is the maximal, Compton cooling-limited CRE Lorentz factor.
The implied CRE-to-CRI energy ratio,
\begin{equation}\label{eq:eta1}
  \eta \equiv \frac{U_{e}}{U_{i}} \simeq 7\times 10^{-4}B^{-3/2}_0 \rho_{-3} \coma
\end{equation}
agrees with the FBs outshining the Galactic disk in high-energy neutrinos (see \S\ref{sec:Intro}), given the $B\gtrsim 1\mu$G fields inferred in the FBs\cite{KeshetEtAl24}.
Note that numerically this $\eta\sim m_e/m_p$, where $m_e$ is the electron mass.

\begin{table}[h!]
  \centering
\begin{tabular}{|cc|cc|cc|}
\hline
RA & Dec & $l$ & $b$ & Reconstruction & $\epsilon$ \\
(deg) & (deg) & (deg) & (deg) & & (TeV) \\
\hline
238.7 & 10.0 & 20.4 & 43.6 & Shower & 191.4 \\
222.5 & -20.9 & -23.1 & 34.1 & Track & 158.4 \\
253.1 & -3.0 & 15.5 & 24.7 & Shower & 148.0 \\
214.9 & -0.2 & -15.7 & 55.4 & Track & 139.0 \\
240.6 & 9.4 & 20.9 & 41.7 & Track & 121.0 \\
240.1 & 0.1 & 10.1 & 37.2 & Track & 117.9 \\
227.1 & 1.4 & 0.7 & 48.3 & Track & 111.5 \\
\hline
\end{tabular}
  \caption{HESE events localized in the northern FB shell\footnote{Not listed is the $\sim102\TeV$ HESE event No.~2, with poor shower localization yielding the two ellipses touching the bottom-left and upper-right edges of the shell in Fig.~\ref{fig:HESE100eROSITA}.}.
  \label{tab:nus}}
\end{table}

\section{Northeastern shell}
\label{sec:NEShell}

The X-rays in  Figs.~\ref{fig:eROSITA2}--\ref{fig:HESE100eROSITA} and in \emph{ROSAT} stacking\cite{KeshetGurwich18}, as well as the \gama-ray\cite{KeshetGurwich17} and radio\cite{KeshetEtAl24} priors, suggest that neutrino emission from the FBs should be limb brightened, with stronger emission from the eastern sectors (\S\ref{sec:Intro}).
These shells subtend a smaller solid angle and indicate a higher coincident nucleon density than the full bubble, implying a higher detection significance and a lower inferred $U_i$ with respect to the above estimates.

Particularly strong X-ray emission from the eastern part of the north FB is seen in Figs.~\ref{fig:eROSITA2}--\ref{fig:eROSITA2Shells}.
The enhanced $-10\dgr<\psi<0$ emission (within $5\dgr$ of the steepest \gama-ray gradient) corresponds to an emission measure jump of order $\Delta\mbox{EM}\simeq0.25\cm^{-6}\pc$ according to the \emph{ROSAT} signal stacked parallel to the edge\cite{KeshetGurwich18}.
For a conservative assumption of uniform density along an $l_{\tiny{\mbox{EM}}}\sim10l_1\kpc$ overlap with the line of sight, the inferred $\rho\simeq 5.7\times 10^{-3}m_p l_1^{-1/2}\cm^{-3}$ is considerably higher than found on average in the high-latitude FBs.

Consider the downstream region within $5\dgr$ of the $b>30\dgr$, $l>0$ \gama-ray edge (dotted curve in Fig.~\ref{fig:eROSITA2Shells}), spanning  $0.0759\sr$, which is a fraction $0.60\%$ of the sky, or $0.66\%$ of the $A$-weighted sky.
The integrated PDF in this region corresponds to $3.15$ neutrinos (out of 164 all-sky), with $2.53$ of them above $100\TeV$ (out of 55 across the sky).
Due to the small size of this region, a random fluctuation can be excluded here only at a modest, $2.5\sigma$ level.
The $U_i\simeq 10^{54}l_1^{1/2}d_1^2\erg$ indicated in this region is smaller by a factor $\sim3$ than anticipated for this region based on Eq.~\eqref{eq:ECRI}, due to the high ambient density.

In order to more robustly quantify the emission from the dense shells, next we adopt a previous\cite{GhoshEtAl26} simple model for ballistic FBs and RBs.

\section{Bubble neutrino model}
\label{sec:mod}

We model the FB edges using an effective axisymmetric 1D stratified model for Galactic bubble evolution\cite{MondalEtAl22, GhoshEtAl26}, which in the ballistic limit indicates a shock radius
\begin{equation}\label{eq:Ballrb}
  r_s \equiv |\vect{r}_s| \simeq z_H\mbox{min}\left[1,\frac{3 \theta_j |z|r}{x^2+y^2+(3-\theta_j)\theta_j z^2} \right]
\end{equation}
terminating the bubble along any ray from the $r=0$, GC origin through inner $\vect{r}=\{x,y,z\}$ Cartesian coordinates,
and adopt a Primakoff-like model\cite{KeshetGurwich18} for their pressure
\begin{equation}\label{eq:P}
  P(\vect{r})\simeq (r/r_s)^3P_d(\vect{r}_s)\simeq (r/r_s)^3 f(z) g(x) P_d
\end{equation}
and mass density
\begin{equation}\label{eq:P}
  \rho(\vect{r})\simeq \frac{r}{r_s}\rho_d(\vect{r}_s)\simeq 4f(z) g(x) \frac{r_0^2r}{r_s^3}\rho_0
\end{equation}
profiles.
Here,
$z_H\simeq 10\kpc$ is the bubble height, $\theta_j\simeq 4\dgr$ is the outflow half-opening angle, the ambient $f(z) g(x) \rho_0(r/r_0)^{-2}$ density profile incorporates a north $f_n=1$ vs. south $f_s=1/2$ bias\cite{KeshetGhosh26} and a $g\simeq 1-x/(6\kpc)$ eastern gradient\cite{GhoshEtAl26}, without which the immediate downstream (subscripts $d$) pressure would be approximately a constant $P_d$ along the entire shock surface\cite{KeshetGurwich18}.

The edges\cite{GhoshEtAl26} and any hadronic $h \propto \rho N_i\propto \rho^{1+\mathfrak{r}} P^\phi$ bubble signals are readily projected along the $l$ line of sight with respect to a solar-system observer at $\{x,y,z\}=\{0,-8.5\kpc,0\}$, for any emissivity determined by constants $\mathfrak{r}$ and $\phi$ approximating the overlap of CRI and target nucleons.
A uniform $P_d$ corresponds to a total outburst energy $U_{tot}\simeq \pi (1+2\theta_j/3) \theta_j z_H^3 P_d/(\Gamma-1)$, where $\Gamma=5/3$ is the adiabatic index; a correction $(f_n+f_s)/2$ is needed to account for possibly different $P_d$ values in each hemisphere.
For a derivation of bubble properties and projections, see Appendix \ref{sec:FBs}.

The anticipated number of FB neutrinos per solid-angle interval $d \Omega$, weighted by the IceCube effective area $A$ and exposure time $t$, becomes
\begin{eqnarray}\label{eq:IceCube}
  \!\!\frac{d n_\nu}{d\Omega} & \simeq & \sum_j p_j \simeq t \int dE_i\, \sum_\alpha A_\alpha\left(\epsilon\simeq f_\alpha E_i\right) \int dl\, \frac{\sigma_i c \, \rho}{4\pi m_p} \frac{dN_i}{dE_i} \, \nonumber \\
  & \simeq & \frac{t\sigma_i c f_\nu}{4\pi m_p}\left(\epsilon_{1}^{-1}-\epsilon_{2}^{-1}\right) A \int dl\, \rho \, C_i \coma
\end{eqnarray}
where $\alpha$ is the flavor and $f_\alpha\simeq f_\nu/3$ is the neutrino energy fraction with respect to the parent CRI.
The measured ${d n_\nu}/{d\Omega}$ excess in a bubble can be used to determine its CRI normalization $C(r_s)$, and thus $P_d\simeq (\Gamma-1)C(r_s)\ln(\gamma_{\Max})/\xi_i$, where $\xi_i$ is the ion acceleration efficiency, eliminated in $U_i\simeq \xi_i U_{tot}(P_d)/2$.

\section{Bubble shell excess}
\label{sec:shells}

\begin{figure}[h]
    \begin{tikzpicture}
    \draw (0, 0) node[inner sep=0, align=center]
    {
        \includegraphics[width=0.47\textwidth,trim={0cm 0cm 0cm 0cm},clip]{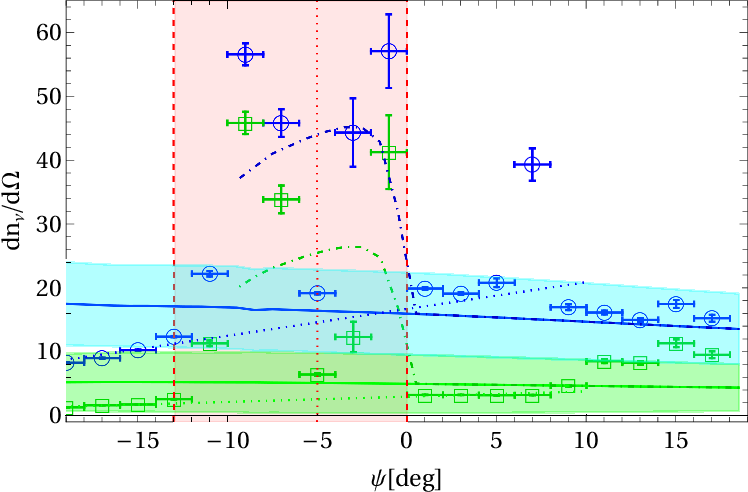}\\[1ex]
        \includegraphics[width=0.47\textwidth,trim={0cm 0cm 0cm 0cm},clip]{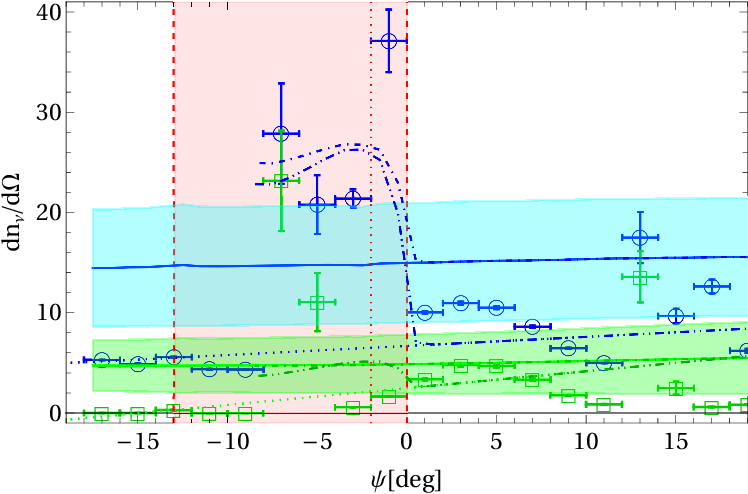}
    };
    \end{tikzpicture}
	\caption{\label{fig:FBsIC}
        All (blue circle) and high energy ($>100\TeV$; green squares) HESE-event PDFs stacked parallel to the north (top panel) and south (bottom) FB edges in $\Delta\psi=2\dgr$ bins.
        The integrated PDFs are elevated inside the X-ray shells identified in Fig.~\ref{fig:eRositaProf} (vertical red dashed lines and shaded region) with respect to the RA-scrambled foreground (solid lines with $1\sigma$ shaded range; see text) or the linear fit $10\dgr$ from the shell (linear fits dotted lines). Best-fit hadronic models are shown on top of these scrambled (dot-dashed) or local (in the south; double-dot dashed) foregrounds.
    }
\end{figure}

The PDFs of HESE events integrated parallel to the FB edges are shown in Fig.~\ref{fig:FBsIC}, in the same method used in Figs.~\ref{fig:RBs} and \ref{fig:eRositaProf} and perviously found to pick-up subtle FB\cite{KeshetGurwich17, KeshetGurwich18, Keshet25PolFB, KeshetEtAl24} and RB\cite{KeshetGhosh26} signals.
The resulting $\psi$ profiles are shown for both north (top panel) and south (bottom) bubbles, along with a simple $h\propto \rho^1 P^1$ model neglecting diffusion (dot-dashed curves).
HESE results are shown for both the full sample (blue circles) and the high, $>100\TeV$ energy sub-sample (green squares), compared to their respective foreground estimates (solid curves with shaded $1\sigma$ range) based on RA-scrambling the data after smoothing over $10\dgr$ scales approximately matching the $\Delta\psi=2\dgr$ bins in terms of solid angle.
These conservative foreground estimates are typically higher than the local foreground actually found around the shell (best linear fits within $10\dgr$ outside the shell shown as dotted curves).

Consider first the north bubble.
For all HESE events, we find an excess throughout the predefined shell, with an integrated $8.2\sigma$ local significance;
with $\chi^2\simeq 88$ in the six shell bins, the RA-scrambled foreground model is rejected at the $8.3\sigma$ level.
The best hadronic fit (dot-dashed curves) corresponds to $U_{i,n}=(3.7\pm0.5)\times 10^{54}\erg$ in this north (subscript $n$) bubble, favored over this foreground model at the $7.7\sigma$ level by the TS-test (see \S\ref{app:TS}).
The high, $>100\TeV$ events show a similar but slightly more significant excess.
Namely, the predefined shell shows an integrated excess of $9.0\sigma$ local significance; with $\chi^2\simeq 125$ in the six shell bins, the foreground model is rejected at the $>10\sigma$ level.
The best hadronic fit corresponds here to $U_{i,n}=(3.1\pm0.4)\times 10^{54}\erg$, favored over the foreground model at the $7.9\sigma$ level.
Outside the shell, the local foreground is seen to be fainter than its RA-scrambled average; switching from scrambled to local foreground would further raise these high confidence levels, with no significant change to hadronic model normalization.

Note that these morphological tests slice the bin-integrated PDFs of only a small number of events, so the conditions for mapping the test statistic to a Gaussian significance are not rigorously fulfilled.
We retain these tests because they exploit the full morphology of the profile relative to the predefined shell, which bulk counting does not.
As a conservative cross-check, treating the bins as independent Poisson variables and computing the joint probability of each shell bin showing the observed excess while neighboring bins remain depleted indicates a $4.7\sigma$ ($4.5\sigma$) excess for all (for $>100\TeV$) energies.
This check is conservative in that the Poisson variance exceeds both the fluctuations expected for bins built from fractional event contributions and the empirical spatial dispersion used by the TS-test to evaluate the structural alignment.

The south FB (subscript $s$; bottom panel of Fig.~\ref{fig:FBsIC}) too shows an excess coincident with the X-ray shell, but at a lower significance.
For all HESE events, we find a modest excess of integrated $1.6\sigma$ local significance in the predefined shell;
with $\chi^2\simeq 19.0$ in the six bins, the RA-scrambled foreground model is rejected at the $2.6\sigma$ level.
The best-fit hadronic model corresponds to $U_{i,s}=(3.1\pm0.9)\times 10^{54}\erg$, favored over this foreground model at the $3.3\sigma$ TS-test level for the four bins we can model in this slightly smaller bubble.
Note that this energy pertains to the assumed factor $f_s=1/2$ reduction in target nucleon density compared to the north.
Similar results are obtained for the $>100\TeV$ statistically-limited sub-sample, but here the inclusion of hadronic emission does not improve the fit above the RA-scrambled foreground.
The inferred excess again strengthens if one replaces the scrambled foreground (solid green curves) with a local fit just outside the shell (dotted green).
The full HESE sample then indicates a strong, $8.2\sigma$ local excess in the shell, rejecting the local foreground at the $9.2\sigma$ level, and favoring the addition of an $U_{i,s}=(5.1\pm0.5)\times 10^{54}\erg$ hadronic component at the $9.4\sigma$ TS-test level.
The $>100\TeV$ sub-sample shows an integrated $3.1\sigma$ local shell excess over the local foreground model, rejecting the latter at the $2.8\sigma$ level.

\begin{figure}[t]
    \begin{tikzpicture}
    \def\dxN{4.5}
    \def\dyN{2.5}
    \def\dxS{-4.4}
    \def\dyS{-2.5}
    \draw (0, 0) node[inner sep=0, align=center]
    {
        \includegraphics[width=0.47\textwidth,trim={0cm 0cm 0cm 0cm},clip]{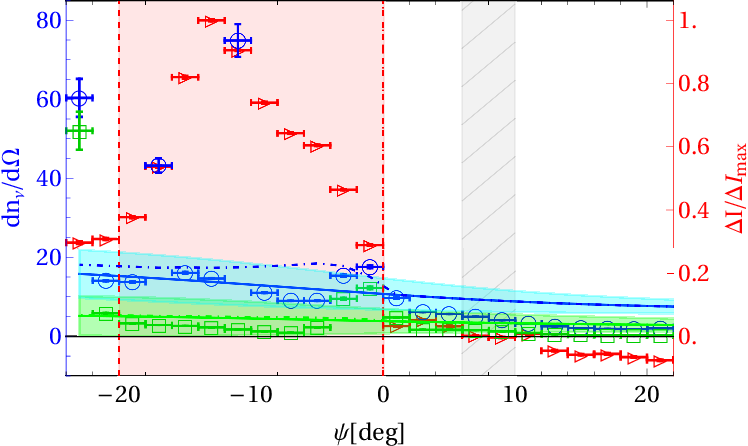}\\[1ex]
        \includegraphics[width=0.47\textwidth,trim={0cm 0cm 0cm 0cm},clip]{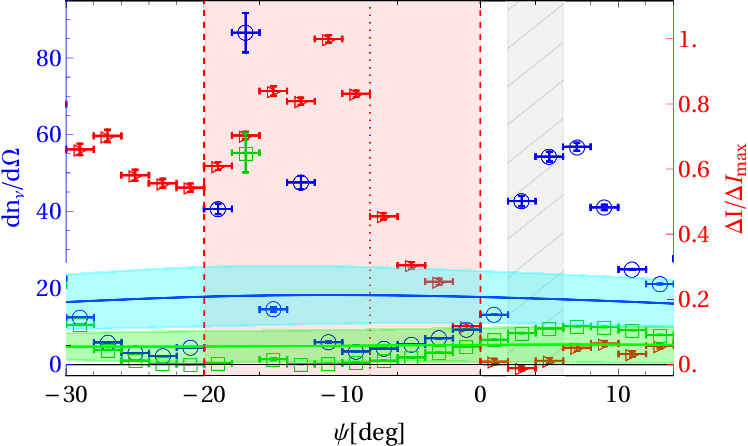}
    };
        \node[text=green!50!black, font=\bfseries] (mytext1) at (-2.9+\dxN,0.2+\dyN) {\scriptsize$391\TeV$ shower};
        \draw[->,green!50!black,thick]  (-3.9+\dxN,0.+\dyN) -- (-4.37+\dxN,-0.55+\dyN);
        \node[text=green!50!black, font=\bfseries] (mytext2) at (2.1+\dxS,1.65+\dyS) {\scriptsize$4.8\PeV$ track};
        \draw[->,green!50!black,thick]  (1.9+\dxS,1.5+\dyS) -- (2.7+\dxS,0.80+\dyS);
    \end{tikzpicture}
	\caption{\label{fig:RBsIC}
        Same as Fig.~\ref{fig:FBsIC} but for the high-latitude eastern RB sector shells identified in Fig.~\ref{fig:RBs}.
        The \emph{eROSITA} $0.6$--$1.0\keV$ profiles of Fig.~\ref{fig:RBs} (normalized to the same hatched region) are over-plotted (red triangles with right axis; taking advantage of the low X-ray foreground).
        Details of the high energy events are provided (green arrows).
    }
\end{figure}
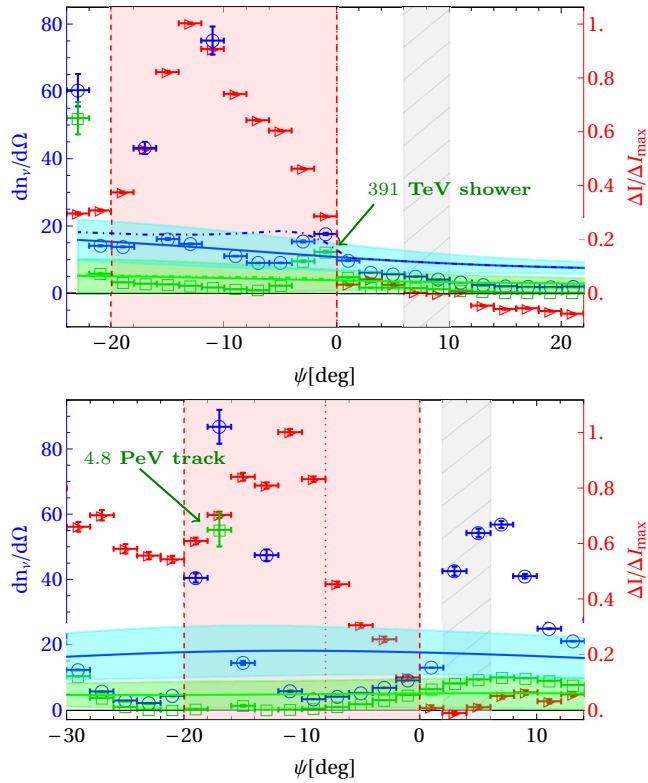

The above analysis is repeated for the high-latitude eastern RB sectors examined in Fig.~\ref{fig:RBs}, as shown in Fig.~\ref{fig:RBsIC}.
In the northeast bubble sector, the full HESE data show an integrated $5.4\sigma$ local excess in the shell, with $\chi^2\simeq 94$ in $10$ bins rejecting the RA-scrambled foreground at the $8.6\sigma$ level; a hadronic ballistic model with $U_{i,n}=(8.6\pm2.9)\times 10^{54} \erg$ is favored over this foreground at the $2.9\sigma$ TS-test level.
Note that the ballistic model used to derive $U_i$ may poorly approximate the  RBs, which appear to be slowing down\cite{GhoshEtAl26}.
The $>100\TeV$ sub-sample shows no excess with respect to the foreground.
In the southeast RB sector, the full HESE data (the $>100\TeV$ sub-sample) show an integrated $1.8\sigma$ ($1.3\sigma$) local excess in the shell, rejecting the foreground at the $8.0\sigma$ ($6.3\sigma$) level; adding a hadronic component to the foreground does not improve the fit.
The shells of these RB sectors harbor two very energetic HESE events, as indicated in the figure.

\section{Summary and discussion}
\label{sec:Summary}

We study the IceCube 12-year HESE data away ($|b|>30\dgr$) from the Galactic plane, finding positive correlations with \emph{Fermi}-LAT sky maps (\S\ref{sec:Corr}).
These correlations are attributed to a significant signal from the FBs, especially the high-density region in the northern bubble, as well as tentative signals from the RBs.
The high-energy, $>100\TeV$ track events in Fig.~\ref{fig:HESE100eROSITA} suffice to indicate a $>4\sigma$ excess in the northern FB (\S\ref{sec:NFB}), which strengthens to $>5\sigma$ when accounting for additional events and competing signals.
X-ray data from \emph{ROSAT} and \emph{eROSITA} trace limb-brightened projected shells in both the FBs (Fig.~\ref{fig:eRositaProf}) and the RBs (Fig.~\ref{fig:RBs}), especially in the north and in eastern sectors, which are used as priors and for bubble modelling (\S\ref{sec:mod}).
Using the RA-scrambled events as a conservative foreground, morphological TS-tests then indicate a $>7\sigma$ ($\sim3\sigma$) excess in the north (south) FB shell, and $>5\sigma$ ($1.8\sigma$) in the north (south) eastern RB sectors.
The locally-measured foreground is typically fainter, corresponding to more significant excess signals (\S\ref{sec:shells}).

Modeling the signal as hadronic emission from inelastic CRI collisions, using either the direct approximation \eqref{eq:ECRI} or the Primakoff-like model in Eqs.~\eqref{eq:Ballrb}--\eqref{eq:IceCube}, indicates that CRI carry a similar $U_i\simeq 10^{54.5}\erg$ (with an uncertainty factor $\sim 3$) energy in each of the FBs and in the northern RB; the neutrino excess in the southern RB is insufficient for such modelling.
These energies are consistent with the total $10^{56\pm1}\erg$ attributed to each of the GC outbursts, which gave rise to the RBs and later to the FBs\cite{GhoshEtAl26}, provided that CRI carry a fraction $\xi_i\sim10\%$ of the thermal energy.
In the FBs, the hard radio-to-microwave emission indicates that CRE carry a fraction $\eta\sim m_e/m_p$ of the CRI energy; see Eqs.~\eqref{eq:ECRE}--\eqref{eq:eta1}.
For such values, one expects comparable high-energy neutrino signals from the Galactic disk and from the FBs (\S\ref{sec:Intro}), so the recent detection of the former is consistent with our detection of the latter.

\acknowledgements
We thank Eli Waxman, Boaz Katz, and Arka Ghosh for helpful discussions.
This research received funding from ISF grant No. 2126/22.

\bibliography{FermiBubbles}

\appendix


\section{Fermi-LAT data}
\label{sec:FermiLAT}

For the \gama-ray data, we examine four channels logarithmically spaced in the $1$--$100\GeV$ range, using the archival Pass-8 LAT data from the Fermi Science Support Center (FSSC)\footnote{See \url{http://fermi.gsfc.nasa.gov/ssc}}, and the Fermi Science Tools (version \texttt{v10r0p5}).
Weekly all-sky files spanning weeks $9$ through $789$ for a total of $781$ weeks ($\sim15$ years) are used, with ULTRACLEANVETO class photon events.
We apply a $90^\circ$ zenith angle cut to avoid CR-generated $\gamma$-rays from the Earth's atmospheric limb, according to
FSSC recommendations, and select good time intervals using the recommended expression \texttt{(DATA\_QUAL==1) and (LAT\_CONFIG==1)}.

\section{Correlation at low resolution}
\label{app:CorrResolution}

\begin{figure}[h!]
    \includegraphics[width=0.49\textwidth,trim={0cm 0cm 0cm 0cm},clip]{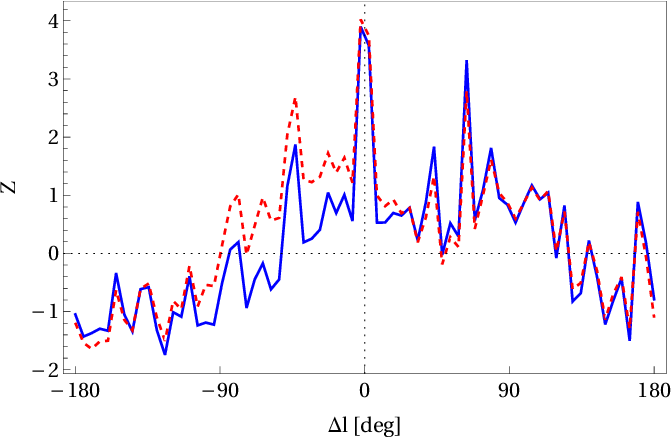}\\
    \vspace{0.3cm}
    \includegraphics[width=0.49\textwidth,trim={0cm 0cm 0cm 0cm},clip]{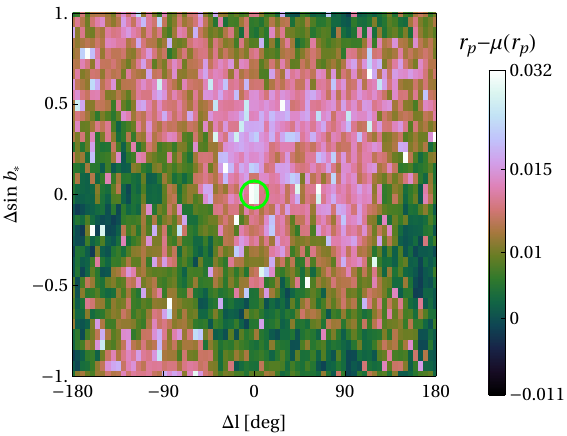}
	\caption{\label{fig:LAT4Corr5deg}
         Same as Fig.~\ref{fig:LAT4Corr} but with $5\dgr$ resolution.
    }
\end{figure}

Figure \ref{fig:LAT4Corr5deg} degrades the resolution of Fig.~\ref{fig:LAT4Corr} to $5\dgr$ in order to demonstrated convergence.

\section{Bubble structure and projection}
\label{sec:FBs}

The ballistic limit of stratified bubble evolution\cite{MondalEtAl22}\coma
\begin{equation}\label{eq:BallRb}
  x_b^2+y_b^2 = R_b^2(z)\simeq z_b^2\left[ 3\left(\frac{z_H}{z_b}-1\right)\theta_j+\theta_j^2 \right] \coma
\end{equation}
indicates that the shock radius satisfies
\begin{equation}\label{eq:BallrbEq}
  r_s \equiv |\vect{r}_s| \simeq \left[3\left(\frac{z_H r}{r_s z}-1\right)\theta_j+\theta_j^2 \right]^{1/2} \frac{r_s z}{R}\coma
\end{equation}
where $R=(x^2+y^2)^{1/2}$ is the cylindrical radius.
Solving for $r_s$ yields
\begin{equation}\label{eq:Ballrb0}
  \!\!\! r_s \equiv |\vect{r}_s| \simeq \mbox{min}\left[\sqrt{R^2+z_H^2},\frac{3 \theta_j z_H |z|r}{R^2+(3-\theta_j)\theta_j z^2} \right] \coma
\end{equation}
or its approximation \eqref{eq:Ballrb}.

The kinetic-to-internal energy ratio is constant along a radial ray,
\begin{equation}\label{eq:Frac}
  \frac{\rho v^2/2}{P/(\Gamma-1)} = \frac{(\Mach^2-1)^2}{(\Mach^2+3)(\Mach^2-1/5)}\simeq 1 \coma
\end{equation}
and is approximately unity for the strong shocks considered.
The total outburst energy is therefore
\begin{eqnarray}
  U_{tot} & \simeq & 4\int_{z>0} \frac{P dV}{\Gamma-1} \simeq
  \frac{4\pi P_0}{3(\Gamma-1)}\int_0^{\pi/2} r_s(\theta)^3\sin{\theta}\,d\theta \nonumber \\
  & \simeq & \frac{\pi (1+2\theta_j/3) \theta_j}{\Gamma-1} z_H^3 P_0 \fin
  \label{eq:TotE}
\end{eqnarray}

The shock front defined by Eq.~\eqref{eq:BallRb} can be projected with respect to the $R_\odot \simeq 8.5$ kpc Galactocentric radius of the
sun by $\{x,\tilde{y},z\}\cdot \bm{\nabla}[x^2+y^2-R_b(z)^2]=0$, to yield\cite{GhoshEtAl26}
\begin{equation}\label{eq:BallProj_l}
  \tan l(z) = \frac{x}{\tilde{y}}
  = \frac{ \sqrt{ [ 12 h - 4\zeta (3 - \theta_j) -
   9 \zeta h^2 \theta_j ] \zeta \theta_j} } {2 - 3 \zeta h \theta_j}
\end{equation}
and
\begin{equation}\label{eq:BallProj_b}
  \frac{1}{\tan^2 b(z)} = \frac{x^2+\tilde{y}^2}{z^2}
  = \frac{1}{\zeta^2} -  (3-\theta_j)\theta_j
\end{equation}
for a range of $0<z<z_H$, where $\zeta\equiv z/R_\odot$ and $h\equiv z_H/R_\odot$ are normalized lengthscales and $\tilde{y}\equiv y+R_\odot$.

We adopt the parameters of Refs.~\citenum{MondalEtAl22} and \citenum{GhoshEtAl26} for modeling the bubbles.
At high latitudes, the ambient density profile is assumed to approach $\rho\simeq 4\times 10^{-4}m_p(r/10\kpc)^{-3/2}\cm^{-3}$.
For the FBs, we adopt $z_H=10\kpc$ and $\theta_j=4\dgr$.
For the RBs, we adopt $z_H=22.9\kpc$ and $\theta_j=4\dgrdot2$ in the north, and $z_H=39.6\kpc$ and $\theta_j=2\dgrdot9$ in the south.

\section{TS-test}
\label{app:TS}

The significance of the hadronic model is inferred from the $\mbox{TS} = \chi^2_- - \chi^2_+$ test, which compares the $\mathsf{n}$ degree-of-freedom $\chi^2$ fit values obtained before ($-$ subscripts) and after ($+$ subscript) adding the model to the foreground; TS then approximately follows \cite{Wilks1938} a chi-squared distribution $\chi_\mathsf{n}^2$ of order $\mathsf{n}\equiv \mathsf{n}_+-\mathsf{n}_-$.

\end{document}